# Subseasonal Predictability and Atmospheric Dynamics of the 2025 Los Angeles Wildfires Revealed by Hybrid Machine-Learning Simulations

Ziming Chen [a], L. Ruby Leung [a], Sandro W. Lubis [a], Jian Lu [b], Wenyu Zhou [a], Sally Wang [a], Ya Wang [c], Yun Qian [a]

[a] *Pacific Northwest National Laboratory, Richland, Washington, USA*

[b] *College of Oceanic and Atmospheric Sciences, Ocean University of China, Qingdao, China*

[c] *State Key Laboratory of Numerical Modeling for Atmospheric Sciences and Geophysical Fluid Dynamics, Institute of Atmospheric Physics, Chinese Academy of Sciences, Beijing, China*

*Corresponding author*: Ziming Chen (ziming.chen@pnnl.gov) & L. Ruby Leung (Ruby.Leung@pnnl.gov)

## ABSTRACT

The January 2025 Los Angeles wildfires (7–31 January 2025) caused widespread fatalities and extensive structural damage. In the preceding weeks, a strong quasi-stationary wave (QSW), characterized by upper-level anticyclonic circulation, persisted over the western United States. Despite its substantial impacts, the dynamics and subseasonal predictability of this circulation pattern remain unexplored.

Here, we investigate the large-scale circulations preceding the fires using a large ensemble of simulations from 1 December 2024 to 31 January 2025, generated by NeuralGCM, a hybrid model combining a dynamical core with machine-learning-based physics. By isolating a 50-member "involution" ensemble constrained solely to observed North American circulation during 5–10 December 2024, we skillfully capture the observed QSW evolution from 11 December 2024 to 7 January 2025 – up to one month in advance of the wildfire event. Sensitivity experiments indicate that atmospheric initial conditions account for ~52% of the QSW anomaly, with Madden-Julian Oscillation (MJO) convection over the tropical western Pacific acting as a key dynamical pathway. The combined MJO and La Niña-related sea surface temperature anomalies prolonged the QSW persistence to 11 days, compared with 6 days in climatology.

These results demonstrate subseasonal predictability of persistent large-scale circulation patterns preceding extreme wildfires up to one month ahead. They also highlight the potential of combining hybrid ML models with physics-guided ensemble sampling to identify windows of opportunity for anticipating such low-likelihood high-impact extreme events on subseasonal timescales.

## SIGNIFICANCE STATEMENT

In January 2025, wildfires across Los Angeles County burned nearly 40,000 acres, destroyed about 16,000 structures, and caused 31 deaths. While local fire weather is often studied, the large-scale atmospheric conditions preconditing such extremes–and their subseasonal predictability–remain poorly understood. Using large-ensemble simulations from NeuralGCM, a hybrid dynamical-machine-learning model, we show that the persistent circulation pattern conducive to the January 2025 fires was predictable about one month in advance. Through hierarchical sensitivity experiments and a physics-guided "involution" ensemble down-selecting, we identify the key dynamical drivers of this event, demonstrating the power of computationally efficient machine learning-based models to anticipate low-likelihood, high-impact extremes on subseasonal timescales.

## CAPSULE

Hybrid machine-learning simulations and physics-guided ensemble sampling reveal an approximately one-month predictive window for the circulation pattern preceding the January 2025 Los Angeles wildfires.

## 1. Introduction

Wildfires exert severe economic and social costs through loss of life, structural damage, and ecosystem degradation (Jones et al. 2022; Qian 2025; Zhang et al. 2025b; Wallace et al. 2026). The January 2025 Los Angeles (LA) wildfires provide a striking example: they burned nearly 40,000 acres, destroyed more than 16,000 structures, caused 31 direct deaths, and were associated with hundreds of indirect fatalities (LAEDC 2025; Rolla et al. 2026). These fires were primed by the persistently fire-conducive environmental conditions across the western United States throughout autumn and winter 2024, which substantially elevated regional wildfire risk and associated losses (Lindsey 2025; Srivastava et al. 2026). These impacts underscore the critical need to understand the large-scale atmospheric patterns that precondition extreme wildfire events and to determine whether such regimes can be anticipated at subseasonal lead times.

Previous studies have identified an important role of large-sale circulation patterns supporting wildfires in California (Dong et al. 2021; Wang et al. 2022, 2026). Progress has also been made in characterizing fire-atmosphere interactions on seasonal to interannual timescales (e.g., Chen et al. 2016; Justino et al. 2020; Liu et al. 2024a,b; Ma et al. 2025), suggesting that wildfire activity can be influenced by remote forcing and, in turn, can exert far-reaching effects on weather and climate in distant regions. Accurate meteorological forecasting and real-time environmental monitoring are essential for effective wildfire preparedness and evacuation. During the January 2025 LA wildfires, NASA's Disasters Program coordinated with response agencies to deliver 22 satellite and airborne data products (Rolla et al. 2026). Continued progress in wildfire monitoring has increasingly relied on high-resolution observations of fuel conditions, soil moisture, and meteorological variables, together with emerging machine-learning methods (Forkel et al. 2017, 2019; Jones et al. 2022; Qian 2025; Ward-Baranyay et al. 2026). While these efforts have improved understanding of wildfire environment, behavior, severity, and monitoring, they do not directly address a complementary question that is central to hazard anticipation: can the large-scale atmospheric circulation pattern that preconditions an extreme wildfire event be predicted in advance? This question is particularly important for the January 2025 LA wildfires, because the event was preceded not simply by anomalous local weather, but by a persistent, large-scale circulation regime over the western United States.

Recent research has also examined the roles of anthropogenic climate change and natural internal variability in shaping the January 2025 LA wildfires environment. Using a multiple linear regression model for the wintertime fire weather index, Srivastava et al. (2026) showed that La Niña-like conditions and the positive phase of the Pacific-North America pattern substantially increased fire weather risk across southern California, whereas anthropogenic climate change played a secondary role. Other studies have also quantified the contributions of climate change and natural variability to the January 2025 LA wildfires from seasonal or longer-term perspectives (Barnes et al. 2025; Cazzaniga and Faranda 2025; Madakumbura et al. 2025; Srivastava et al. 2026). Key questions remain unresolved. In particular, the dynamical origins of the persistent large-scale circulation pattern that organized the fire weather conditions, and its potential predictability, are poorly understood. It also remains unclear whether the temporal evolution of such fire-conducive circulation anomalies can be predicted to support a process-oriented attribution framework, such as the storyline approach.

These questions are especially relevant because, in the weeks preceding the LA wildfires, a strong quasi-stationary wave persisted over the western United States (Fig. 1A). This circulation anomaly was characterized by upper-level anticyclonic conditions favorable for sustaining hot, dry, and offshore-flow-prone environments over Southern California. The persistent quasi-stationary wave raises a fundamental question: can we identify windows of opportunity for anticipating the large-scale patterns favoring extreme wildfires by combining ML-based models with physics-guided ensemble sampling on subseasonal timescales?

To address this gap, we develop a hierarchical involution framework to improve the subseasonal prediction of the large-scale condition anomalies associated with the January 2025 LA wildfires. We apply this framework to the circulation evolution during December 2024 and address two key questions:

1. Predictability: Can a physics-guided "involution" ensemble using a hybrid-ML model skillfully predict the temporal evolution of the pre-fire QSW pattern at subseasonal lead times (up to one month);

2. Dynamics: What physical mechanisms–including atmospheric initial conditions, tropical Pacific sea surface temperatures, and dynamical pathways-govern the persistence and predictability for this circulation regime?

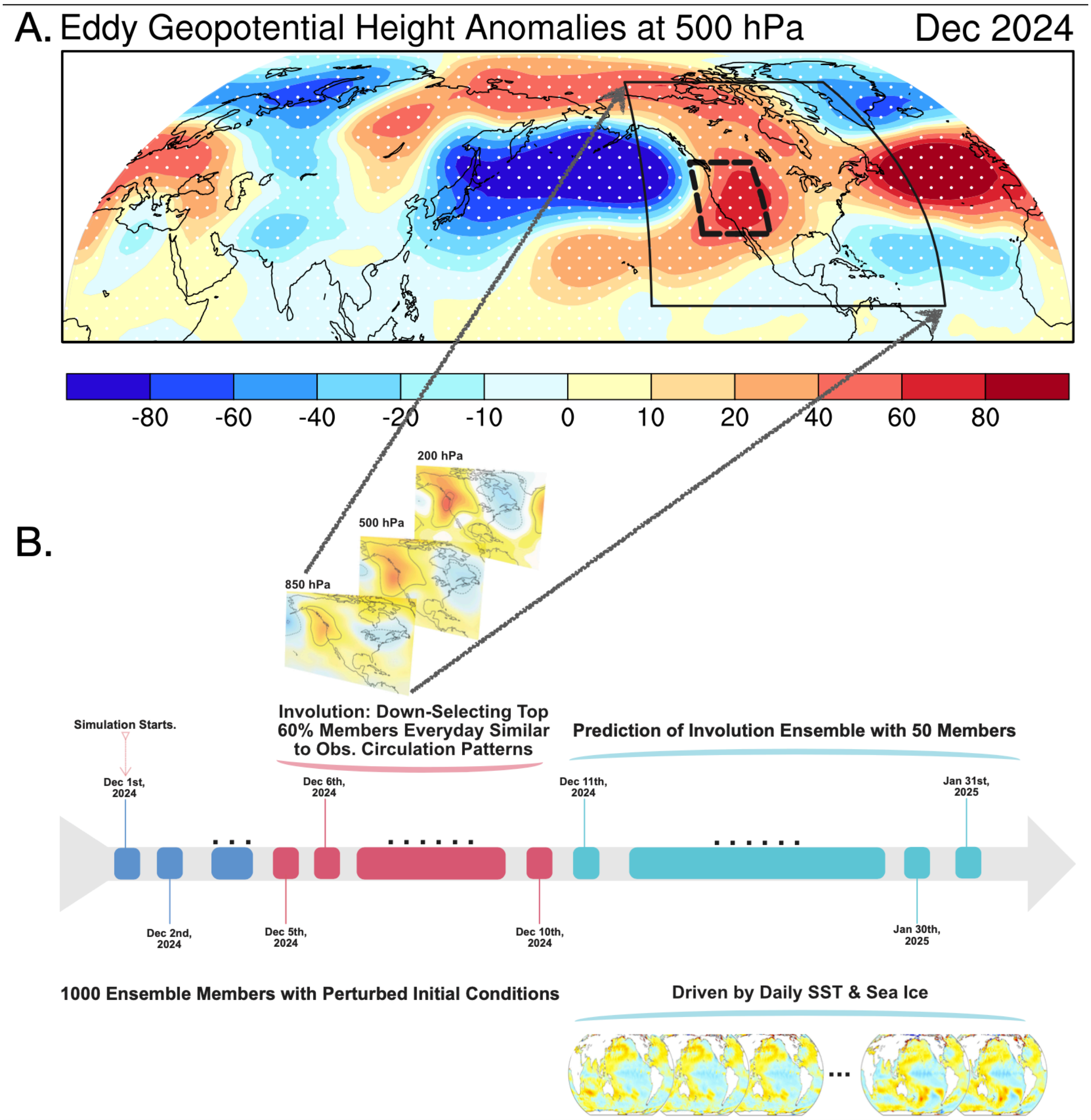

**Figure 1. Large-scale circulation pattern preceding the January 2025 LA wildfires and workflow of the involution ensemble.** (A) ERA5 500-hPa eddy geopotential height anomalies in December 2024 relative to the 1979–2024 climatology. Stippling indicates anomalies that are statistically significant at the 10% level. The thin box marks the North America domain used in the involution algorithm (10ºN~75ºN, 45ºW~150ºW). The thick dashed box indicates the western United States domain (30ºN~50ºN, 105ºW~130ºW). (B) Schematic workflow of the involution ensemble based on ERA5 two-dimensional circulation patterns at the lower, middle, and upper tropospheric levels. The 1000-member ensemble simulations are initialized on 1 December 2024 and integrated through 31 January 2025, using observed sea surface temperature and sea ice concentration as boundary conditions. Individual ensemble members differ from the ERA5 atmospheric initial conditions only by micro perturbations. The perturbations were incorporated into all the fields of initial conditions. During 5–10 December 2024, the root-mean-square error (RMSE) of the anomalous circulation

relative to the ERA5 target field over North America is calculated for each ensemble member, and the top 60% of members with smallest RMSEs are retained each day. The final involution ensemble consists of 50 members and is used to predict the atmospheric conditions from 11 December 2024 to 31 January 2025.

## 2. Conceptual Framework of Hierarchical Involution Simulations

### a. Observational and reanalysis datasets

We use observational and reanalysis datasets both to provide model boundary conditions and to evaluate simulated circulation anomalies. Large-scale atmospheric circulation is characterized using the fifth-generation European Centre for Medium-Range Weather Forecasts reanalysis (ERA5), which has a native horizontal resolution of 0.25°×0.25° and 137 vertical levels extending from the surface to 80 km, and spans 1950 to the present (Hersbach 2000). To identify La Niña-related SST anomalies, we use the Extended Reconstructed Sea Surface Temperatures Version 5 (ERSST v5), which covers the period from 1854 to the present (Huang et al. 2017). To track the temporal phase evolution of the Madden-Julian Oscillation (MJO), we use the real-time multivariate MJO (RMM) index from the Australian Bureau of Meteorology (https://www.bom.gov.au/clim_data/IDCKGEM000/rmm.74toRealtime.txt). The daily RMM1 and RMM2 indices are calculated following Wheeler and Hendon (2004), with the revised procedure of Gottschalck et al. (2010) applied from 2014 onward. Additional details and processing details are provided in the Appendix.

### b. Large ensemble from hybrid ML-based model

We use NeuralGCM, a hybrid ML-based model that combines a differentiable atmospheric dynamical core with ML-based physical parameterizations (Kochkov et al. 2024; Yuval et al. 2026). Specifically, we employ the stochastic version of NeuralGCM trained on ERA5 atmospheric fields and Integrated Multi-satellitE Retrievals for Global Precipitation Measurement (IMERG) version 7 precipitation during 2001–2018, at 2.8° horizontal resolution and 37 vertical $\sigma$-levels (Yuval et al. 2026). Given prescribed SST and sea ice concentration, NeuralGCM predicts the atmospheric states at the next time step and can generate both deterministic and ensemble forecasts, with skill comparable to or exceeding that of pure ML- and conventional physics-based models (Kochkov et al. 2024).

We perform a 1000-member ensemble simulation in which all members are initialized from ERA5 atmospheric states at 0000 UTC 1 December 2024 and integrated hourly through 2300 UTC 31 January 2025. The ensemble members differ only through micro perturbations to the atmospheric initial conditions. Daily SST and sea ice concentration from ERA5 are prescribed as lower boundary conditions throughout the simulations.

The anomalies of circulation patterns are relative to the long-term free-running AMIP transient simulations with NeuralGCM, driven by daily SST and sea ice in 1979–2020, from (Chen et al. 2026). Three ensemble members with different initial conditions are used.

**c. Involution ensemble workflow**

To identify ensemble members that best capture the observed pre-fire circulation evolution, we develop a hierarchical involution framework that constrains large-scale three-dimension dynamical circulation anomalies over North America, during the month preceding the January 2025 wildfires (Fig. 1B). The method begins with the full 1000-member ensemble and iteratively retains the members that most closely match the observed circulation.

Starting on 5 December 2024, we evaluate each ensemble member against ERA5 using area-weighted root-mean-square error (RMSE) over North America for circulation anomalies at 850, 500, and 200 hPa. The target fields include zonal and meridional wind components at these three levels. On each day, only the top 60% of ensemble members with smallest RMSEs are retained. This iterative refinement is repeated daily from 5 to 10 December 2024. The final involution ensemble consists of 50 members and is then used to predict the atmospheric conditions from 11 December 2024 to 31 January 2025.

Similar approaches have been used to constrain dynamical conditions during extreme events (Duan et al. 2025), and recent studies have also explored optimizing initial conditions for ML-based forecasts using backpropagation and gradient descent (Vonich and Hakim 2024, 2026; Xiang et al. 2026). Despite that, those methods generally require knowledge of the future circulation or specifically extreme event trajectory during model training or optimization, which may limit their general applicability and obscure the physical pathways through which predictability is enhanced (Selz and Craig 2026). In contrast, our framework constrains only the observed circulation over a limit region and time window before the event. This design provides a practical test of whether observed large-scale atmospheric conditions over North America, about one month before the fires, contain usable subseasonal predictive information for the circulation pattern that later favored the January 2025 LA wildfires.

## 3. Improved Predictability by the Involution Ensemble

**a. Involution ensemble captures the temporal evolution of the quasi-stationary wave**

Before the January 2025 wildfires, a pronounced quasi-stationary anticyclonic circulation developed over the western United States, as indicated by positive eddy geopotential height anomalies (Fig. 1A). This persistent quasi-stationary wave favored large-scale subsidence, reduced cloud cover, and enhanced downward shortwave radiation, thereby creating atmospheric conditions conducive to wildfire development (Kautz et al. 2022; Chen et al. 2023, 2025; Lubis et al. 2024; Lindsey 2025).

To reproduce this quasi-stationary anticyclonic circulation, we construct a 50-member involution ensemble from the 1000-member NeuralGCM large ensemble by constraining the circulation evolution over North America toward observations (Fig. 1B). Figure 2 compares the involution ensemble with ERA5, the full large ensemble, and a control simulation. The involution ensemble captures both the spatial pattern and temporal evolution of the quasi-stationary wave during December 2024. In particular, it reproduces the significantly positive eddy geopotential height anomalies over the western United States, although with some spatial displacement relative to ERA5 (Fig. 2A). Importantly, the involution ensemble also reproduces large-scale circulation anomalies outside the constrained North American domain. Over the entire Northern Hemisphere, the simulated eddy geopotential height anomaly pattern resembles that in ERA5, with a pattern correlation coefficient (PCC) of 0.61 and a normalized root-mean-square error (NRMSE) of less than 0.8 times the observed standard deviation (Fig. 2A). By contrast, the control simulation, initiated at 0000 UTC 1 December 2024 and forced by daily ERA5 SST and sea ice concentration, produces only a weak and statistically insignificant positive anomaly over the western United States, with a substantial location error (Fig. S1A). More importantly, it fails to reproduce the observed large-scale Northern Hemisphere pattern, with a much lower PCC (0.09) and a larger NRMSE (1.35) than the involution ensemble.

We further assess predicted skills relative to the full 1000-member NeuralGCM ensemble. To enable a fair comparison with the 50-member involution ensemble, we randomly select 50 members from the large ensemble and repeat this sampling 100 times to construct a random ensemble benchmark. This random ensemble shows limited skill in reproducing the anomalous eddy geopotential height pattern of December 2024, with a mean PCC of 0.24 and a mean NRMSE of 1.18 (Fig. S1B), again substantially worse than the involution ensemble.

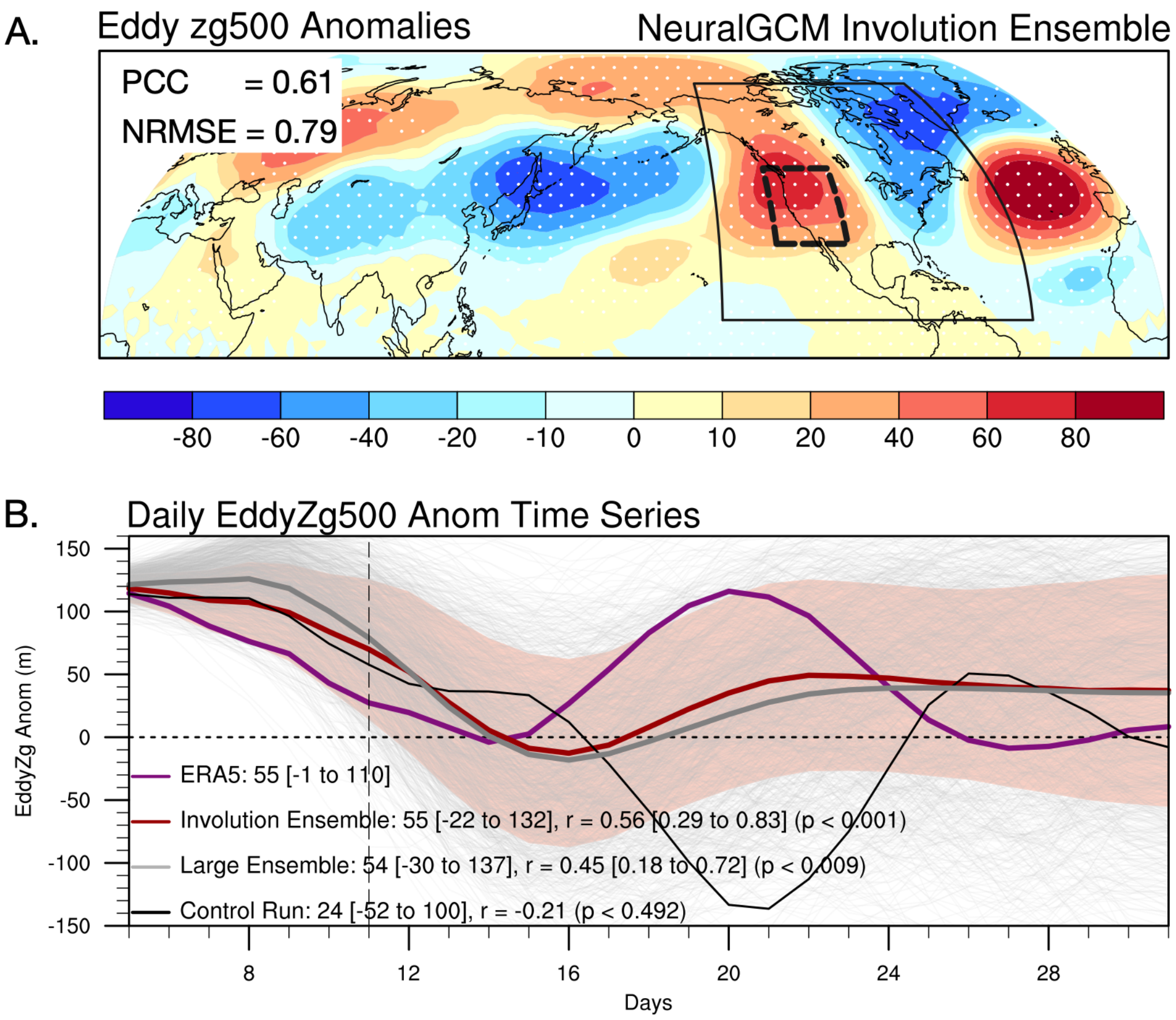

**Figure 2. NeuralGCM involution simulations to the temporal evolution of the quasi-stationary wave.** (A) December 2024 anomalous 500-hPa eddy geopotential height (eddy zg500) in the involution ensemble. Anomalies are defined relative to the climatological mean from NeuralGCM historical AMIP simulations in 1979–2020, forced by observed sea surface temperature and sea ice concentration. Stippling indicates grid points where at least 75% of the 50 ensemble members agree on the sign of the anomaly. The pattern correlation coefficients (PCCs) and normalized root-mean-square errors (NRMSE) between ERA5 and NeuralGCM eddy zg500 anomalies over the Northern Hemisphere are shown in the upper-left corner. NRMSE is normalized by the standard deviation of ERA5 eddy zg500 anomalies. (B) Area-average daily eddy zg500 over the western United States (dashed black box in A) from ERA5 (purple), the NeuralGCM involution ensemble (red, IEM), the NeuralGCM large ensembles (gray), and the control simulation (black). Red shading denotes $\pm 1$ inter-member standard deviation across the 50 involution ensemble members. The values shown next to the legends indicate the time-mean eddy zg500 anomaly and ite temporal correlation with ERA5. For ERA5 and control simulation, the spread in the legend represents the $\pm 1$ temporal standard deviation, while those for NeuralGCM ensemble simulations denote $\pm 1$ inter-member standard deviation. The vertical dashed line marks the start of the prediction period. Units: m

The involution ensemble also better captures the observed temporal evolution of the quasi-stationary wave. In ERA5, a persistent positive 500-hPa eddy geopotential height anomaly is evident over the western United States beginning on 11 December 2024, with a time mean of 55 m during prediction period (purple curve in Fig. 2B). The involution ensemble closely tracks this evolution, with the same time mean of 55 m and a temporal correlation of 0.56 with ERA5 ($p < 0.001$). The 1000-member large ensemble also reproduces a similar anomaly but exhibits a much larger inter-member spread and a lower temporal correlation with ERA5 (0.46, $p < 0.009$). Both ensemble forecasts outperform the control simulation, which loses predictive skill after 17 December 2024. In the following sections, we therefore focus on the source of predictability in the involution and large ensembles.

**b. Atmospheric initial conditions dominate the occurrence of the quasi-stationary wave**

To investigate the source of predictability for the quasi-stationary wave, we design two sensitivity experiments based on the 50-member involution ensemble: AMIP-IC and AMIP-ClimSST. To isolate the role of atmospheric internal variability, the AMIP-IC experiment uses atmospheric initial conditions randomly sampled from 1979 to 2023 while forced by observed SST and sea ice concentration from 1st December 2024 to 31st January 2025. Conversely, to quantify the role of ocean boundary forcing, the AMIP-ClimSST experiment retains the realistic atmospheric initial conditions on 1 December 2024 but replaces the observed SST and sea ice concentration with their climatological values. Figure 3 summarizes the relative contributions of atmospheric initial conditions and ocean boundary forcings. To minimize the direct influence of initialization, we focus on the period beginning 11 December 2024. Relative to all-forcing involution ensemble (Fig. 3A), atmospheric initial conditions emerge as the dominant source of quasi-stationary wave variability. When the realistic atmospheric initial conditions are removed in AMIP-IC, the significant positive eddy geopotential height anomalies over the western United States largely disappear (Fig. 3B). In contrast, in AMIP-ClimSST, significant positive anomalies remain over the western United States even under climatological SST and sea ice forcing (Fig. 3C), although their location differs somewhat from that in the all-forcing involution ensemble and in ERA5 (Figs. 3A & 1A). These results indicate that the realistic atmospheric initial conditions are essential for generating the quasi-stationary wave, whereas ocean boundary forcing plays a secondary role.

We further quantify the relative role of atmospheric variability and ocean boundary forcings using the area-mean difference in anomalous 500-hPa eddy geopotential height over the western United States between the all-forcing involution ensemble and the sensitivity runs (Fig. 3D). Atmospheric initial conditions account for approximately 52% of the total quasi-stationary wave variation, whereas ocean boundary forcings contributes about 12%. The remaining 34% is represented by a residual "others" term, which may include land-surface influences, nonlinear interactions, and unresolved residual contributions. Overall, these results show that atmospheric initial conditions dominate the occurrence of the quasi-stationary wave, while observed SST and sea ice anomalies provide an additional but smaller contribution.

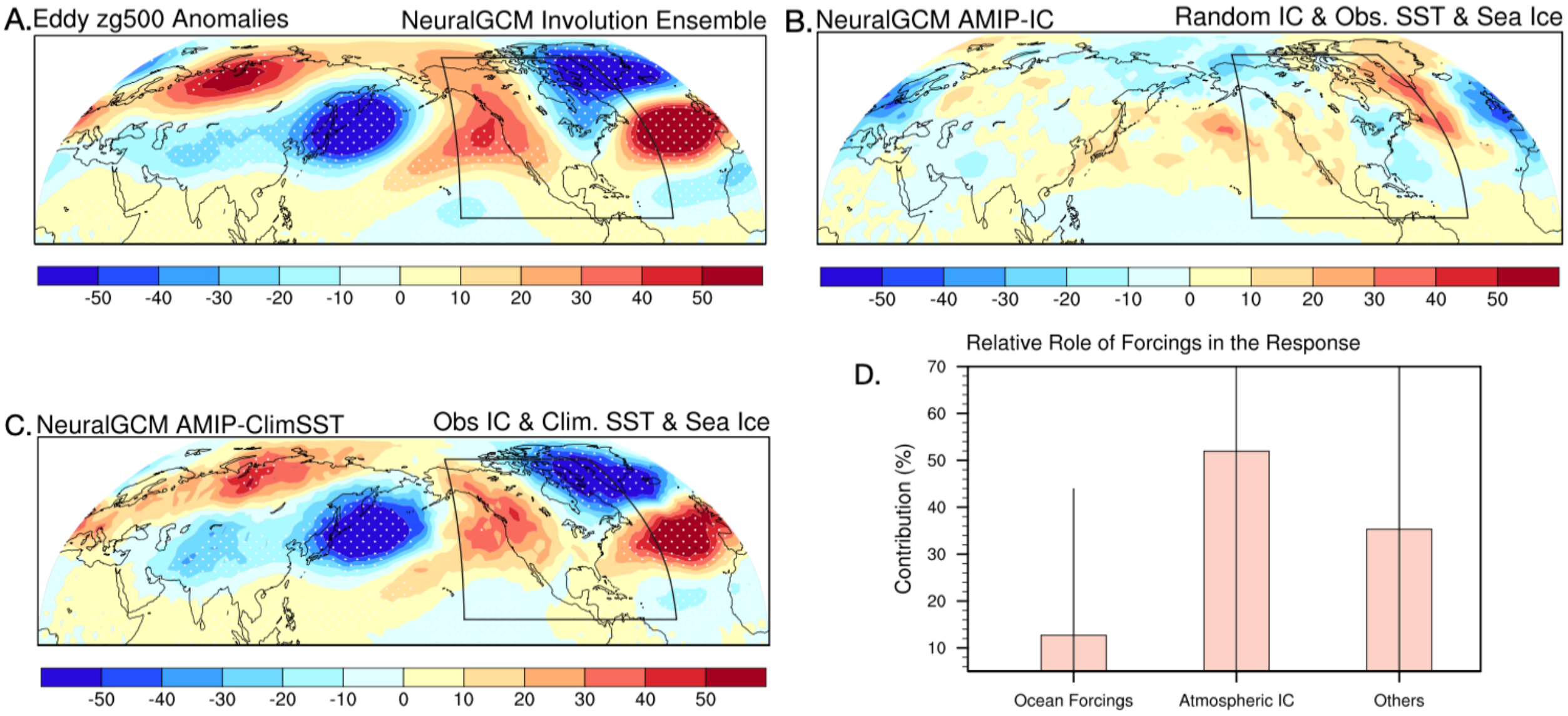

**Figure 3. Contribution of ocean boundary forcing and atmospheric initial conditions to the quasi-stationary wave.** (A) Anomalous 500-hPa eddy geopotential height (eddy zg500) averaged over 11 December 2024 – 7 January 2025 in the NeuralGCM involution ensemble. Anomalies are defined relative to NeuralGCM historical AMIP simulations during 1979–2020. Stippling indicates grid point where more than 75% of the 50 ensemble members agree on the sign of the anomaly. (B, C) same as in (A), but for the NeuralGCM AMIP-IC and AMIP-ClimSST sensitivity experiments, respectively. In AMIP-IC, atmospheric initial conditions are randomly sampled from 1979–2023, while forced by observed SST and sea ice concentration. In AMIP-ClimSST, observed atmospheric initial conditions are retained but forced by climatological mean SST and sea ice concentration. Differences from the all-forcing involution ensemble therefore isolate the effects of atmospheric initial conditions and ocean boundary forcing, respectively. (D) Relative contributions of ocean boundary forcings, atmospheric initial conditions, and other effects to anomalous eddy zg500 over the western United States. Contribution are calculated from the percentage differences between all-forcing involution ensemble and the sensitivity experiments, normalized by the anomalous eddy zg500 in the all-forcing involution ensemble. Bar shows the ensemble mean across the 50 members, and vertical lines denote $\pm 1$ inter-member standard deviation.

What atmospheric processes associated with the initial conditions favor the development of the quasi-stationary wave over the western United States? Previous studies have shown that on S2S timescales, MJO-related convective anomalies over the tropical western Pacific can excite a Pacific-North America (PNA)-like wave train. This wave train is characterized by strong eddy geopotential height anomalies near the exit of the subtropical jet over the northeast Pacific, where quasi-stationary waves over the western United States can extract energy from the zonally asymmetric climatological flow (Wallace and Gutzler 1981; Simmons et al. 1983; Ting and Yu 1998; Mori and Watanabe 2008; Zhou et al. 2020). In particular, previous studies have reported positive 500-hPa geopotential height anomalies and enhanced winter blocking frequency over the western United States during MJO phases 6 and 7 (Henderson et al. 2016, 2017), favoring large-scale subsidence and a fire-friendly environment. Consistent with this mechanism, the observed RMM index indicates active MJO conditions in phases 5–7 during December 2024, before the wildfire outbreak in January 2025 (Fig. 4A).

Accounting for the delayed midlatitude response to tropical convection mediated by Rossby wave propagation (Henderson et al. 2016, 2017; Stan et al. 2017; Lubis et al. 2024; Chen et al. 2025), active MJO phase 6 roughly two pentads before the wildfire would be expected to favor a downstream anticyclonic anomaly over North America. To test this hypothesis, we construct lagged composites of anomalous 500-hPa eddy geopotential height from day 0 to day +9 following MJO phase 6 (Fig. 4D). The composite shows a weak anticyclonic anomaly over the tropical western North Pacific, followed by a cyclonic anomaly over the broader North Pacific, and an anticyclonic anomaly centered over northwestern North America that extends into the Pacific Northwest. Thus, the climatological MJO teleconnection pattern over the past 46 years places the downstream anticyclonic center somewhat northwest of the positive eddy geopotential height anomalies observed over the western United States in December 2024 (Figs. 2A & 4D).

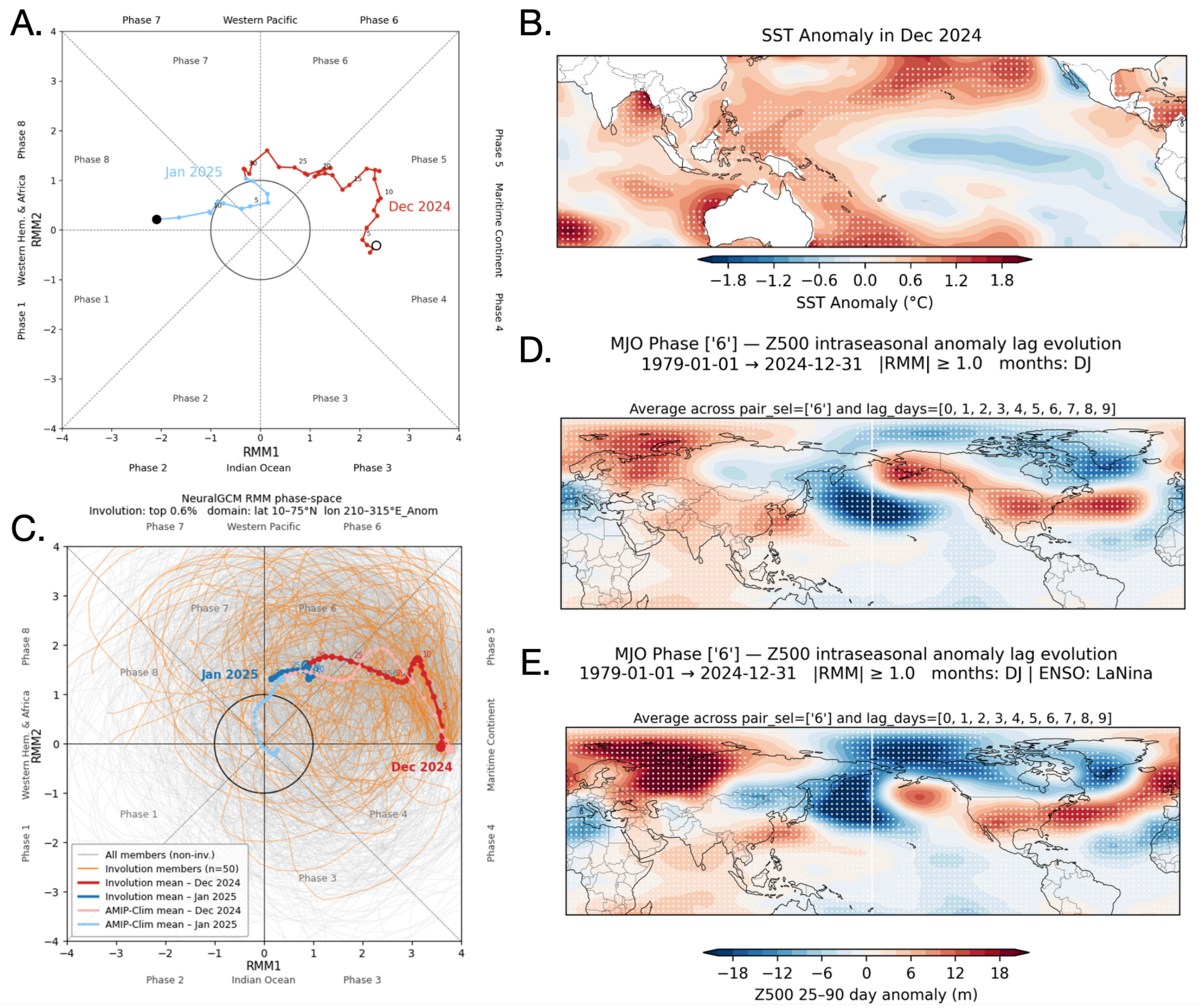

**Figure 4. Observed climate variability during December 2024 – January 2025 and its representation in NeuralGCM.** (A) Observed real-time multivariate (RMM) index with MJO phase-space evolution from 1 December 2024 to 12 January 2025. (B) Monthly SST anomalies in December 2024 relative to 1979–2020. (C) Simulated RMM phase evolution in NeuralGCM. Gray lines show the evolution in the 1000-member large ensemble, and orange lines show the 50 involution ensemble members. Red and blue dot lines denote the inter-member mean RMM evolution of the involution ensemble in December 2024 and January 2025, respectively. Light red and blue lines show the corresponding evolution in the AMIP-ClimSST experiment. (D) Lagged composite of anomalous 500-hPa eddy geopotential height over days 0 to +9 following MJO phase 6 during December–January for 1979–2024. (E) same as in (D), but for La Niña years only. Stippling indicates anomalies that are statistically significant at the 10% level.

What explains this location difference between the climatological MJO teleconnection and the observed circulation anomaly in December 2024? Previous studies have shown that equatorial Pacific SST anomalies can modify intraseasonal Rossby wave propagation during ENSO events (Roundy et al. 2010; Henderson and Maloney 2018; Lee et al. 2019; Fernandes and Grimm 2023). A weak La Niña developed in December 2024 (Fig. 4B; (WMO 2025;

Srivastava et al. 2026). La Niña-related negative SST anomalies tend to strengthen the mean meridional geopotential height gradient over the North Pacific, and thereby intensiy the subtropical westerly jet. Besides, compared with neutral and El Niño years, the Rossby waveguide – represented by maximum stationary wavenumber – becomes broader over the northeast North Pacific during La Niña. Under these conditions, MJO-induced convection over the tropical western Pacific can generate a wave train that propagates more zonally within the jet, anchoring positive geopotential height anomalies farther south over western North America and increasing the likelihood of a dry, hot background state over the western United States (Henderson and Maloney 2018; Wang et al. 2024). To test this interpretation, we conduct lagged composites for MJO phase 6 using only La Niña years (Fig. 4E). Relative to the all-year composite, the center of the positive eddy geopotential height shifts southward toward western North America, with significant positive anomalies appearing directly over the western United States. This result also helps explain the location difference between the all-forcing involution ensemble and AMIP-ClimSST: without the La Niña SST anomalies, the simulated positive height anomaly tends to be displaced relative to the observed pattern (Figs. 3A & 3C).

To further quantify the likelihood that MJO-related convection favors the development of a strong quasi-stationary wave, we calculate the conditional frequency of strong quasi-stationary wave events given active MJO conditions in observations (Fig. 5; see Appendix). To increase the sample size, we combine consecutive MJO phases with similar convective structures into four phase pairs: 8–1, 2–3, 4–5, and 6–7. Strong quasi-stationary wave days are defined as days when the area-mean eddy geopotential height anomaly over the western United States exceeds one temporal standard deviation during 1979–2024. Among all phase pairs and all active-MJO days, the highest conditional probability of strong quasi-stationary wave occurrence is found during MJO phases 6–7, reaching 26%~34% over lead times of 0–2 pentads (Fig. 5A). Conversely, when conditioning on strong quasi-stationary wave occurrence and looking backward in time, MJO phases 6–7 again occur most frequently at a lead of about 15 days, with frequencies of (28%~35% (Fig. 5B).

We further quantify this relationship using logistic regression to estimate the probability that strong quasi-stationary wave days are associated with specific MJO phase pairs (Fig. 5C; see Appendix). Logistic regression is well suited for this binary classification problem and provides an estimate of the odds ratio for strong quasi-stationary wave occurrence during each MJO phase pair. An odds ration greater than 1 indicates that strong quasi-stationary wave events are more likely during MJO phases 6–7 than during other active MJO phases (Fig. S2A). Over 1979–2024, strong quasi-stationary wave occur on 31% (29%~33%) of days at lag of nine days following active MJO phases 6–7, which is about 35% higher than the corresponding probability for all other active phases (23% [21%~25%]; Fig. 5C left). Excluding La Niña years yields a similar conclusion, although the duration of enhanced probability is shorter than in the all-year sample (Fig. 5C right & Fig. S2B). Therefore, active MJO events in phases 6–7 are associated with the subsequent occurrence of strong quasi-stationary wave conditions over the western United States.

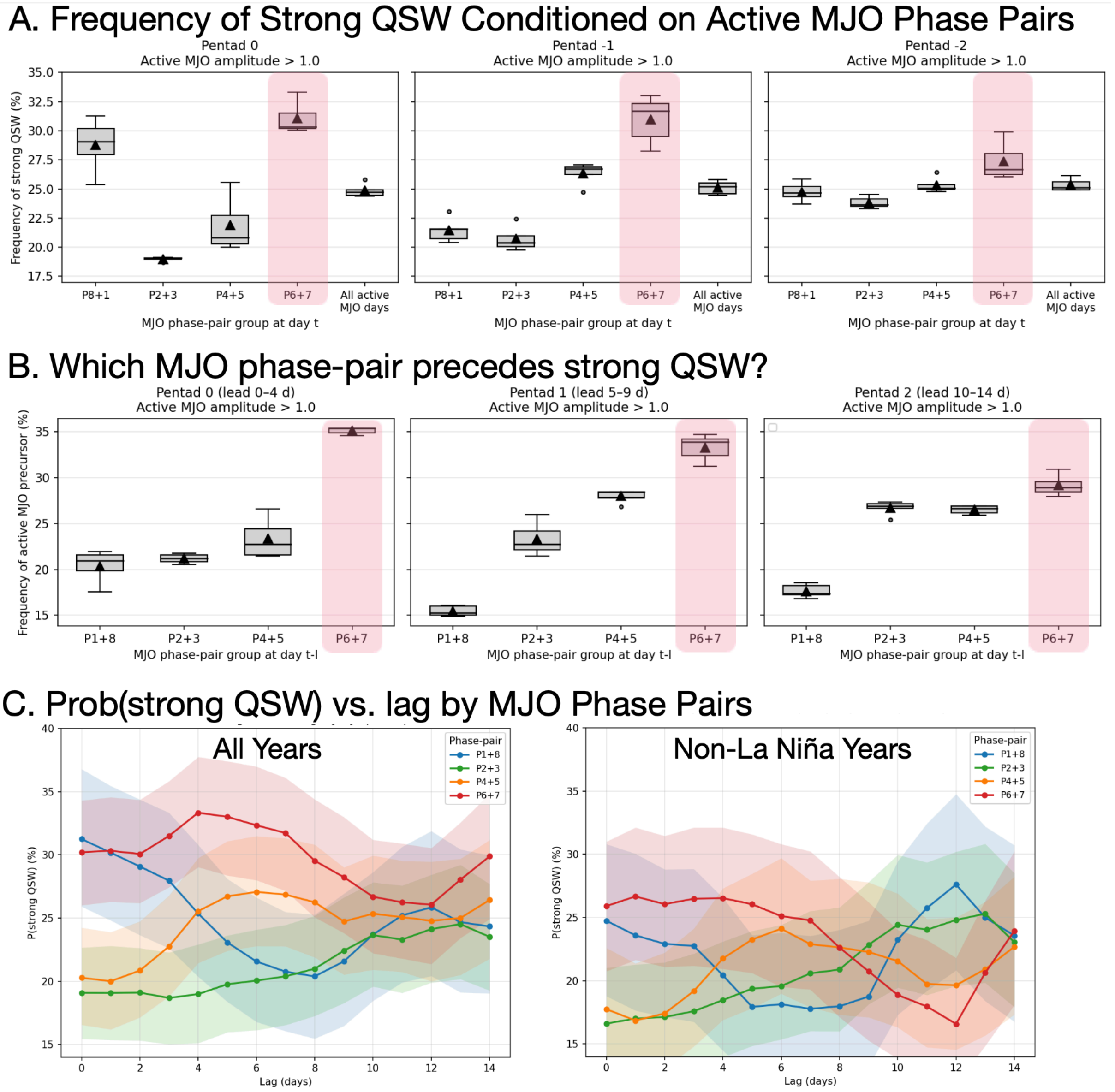

**Figure 5. Conditional frequency and logistic regression analyses highlight strong quasi-stationary wave (QSW) occurrence following active MJO phase 6–7.** (A) Frequency of strong quasi-stationary wave events conditioned on four active-MJO phase pairs and on all active-MJO days, averaged over lead times of

0, 1, and 2 pentads. Boxplots show the interquartile range (box), the 5th–95th percentile range (whiskers), and min/max values (dots). Triangles mark the mean. Pink shading highlights the phase pair with the highest frequency. (B) same as in (A), but showing the frequency of active MJO phase pairs preceding strong quasi-stationary wave events by 0, 1, and 2 pentads. (C) Probability of strong quasi-stationary wave occurrence associated with each MJO phase pair from logistic regression ($P(QSW^{*}(t+l) \geq 0 \mid I_g(t) = 1)$) for all years during 1979–2024 and for non-La Niña years only (see Appendix). Shading indicates the 95% confidence intervals estimated from bootstrap resampling.

Having identified the dynamical pathway linking tropical convection to the quasi-stationary wave, we next ask whether NeuralGCM can reproduce these features. Recent studies have shown the NeuralGCM simulates MJO teleconnections with relatively high fidelity (Diao and Barnes 2025; Peings et al. 2025). It is therefore plausible that the model can capture enhanced predictability of the quasi-stationary wave when the simulated MJO phase evolution resembles the observed ones. Figure 4C shows the RMM phase evolution in NeuralGCM. Although the full 1000-member large ensemble exhibits a broad distribution across all eight MJO phases (gray lines in Fig. 4C), the 50-member involution ensemble is concentrated in phase 5–7 (orange lines in Fig. 4C), with the ensemble-mean RMM index reaching phase 6 in late December 2024 (red and blue dot lines in Fig. 4C). During the prediction period beginning 11th December 2024, about 37% of days in the large ensemble within MJO phases 6–7. In contrast, the involution ensemble yields a slightly higher fraction, about 41%, despite having only 50 members. This result suggests that the involution workflow can effectively identify members with enhanced subseasonal predictive skill from a much large ensemble. More broadly, the combination of computationally efficient ML-based models and physics-guided ensemble sampling highlights a promising pathway for identifying windows of opportunity to anticipate low-likelihood, high-impact extreme events on subseasonal timescales.

### c. La Niña SST anomalies prolong quasi-stationary wave persistence over the western United States

An interesting result is that the AMIP-ClimSST experiment, which uses realistic atmospheric initial conditions but climatological SST, still reproduces the obesrved MJO phase evolution during December 2024 (light red and blue lines in Fig. 4C), broadly similar to that in the involution ensemble (dark red and blue lines in Fig. 4C). In contrast, the active MJO event decays in January 2025 in AMIP-ClimSST, indicating a substantially shorter duration than the involution ensemble, which includes the observed La Niña SST forcings.

Previous studies have shown that, during La Niña years, MJO-related convective anomalies tend to remain over the tropical western Pacific longer than during neutral or El Niño years, thereby favoring a more zonally extended wave train and a persistent positive geopotential height anomaly over western North America (Henderson et al. 2017; Henderson and Maloney 2018). This suggests that La Niña SST anomalies may not be the primary driver of quasi-stationary wave onset, but they can contribute to produce a longer persistence by anchoring the MJO-related anticyclonic anomaly over the western United States.

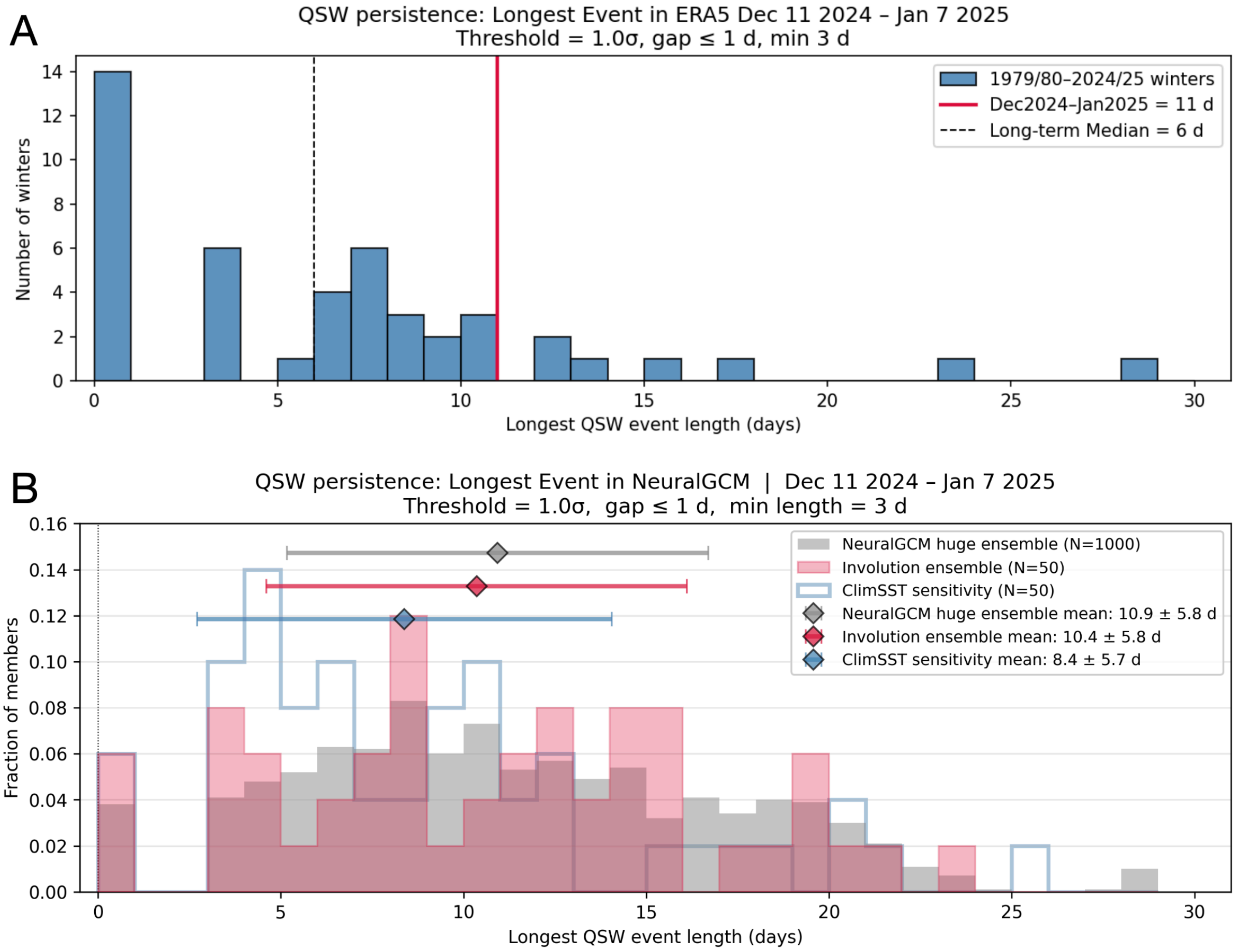

**Figure 6. Persistence of the quasi-stationary wave (QSW) over the western United States.** (A) Quasi-stationary wave persistence during 11 December 2024 – 7 January 2025 (red), compared with the distribution during 1979–2024 in ERA5 (blue bars). Quasi-stationary wave persistence is defined as the number of consecutive days on which the area-mean eddy geopotential height anomaly over the western United States exceeds one temporal standard deviation. Events must persist for at least three days, with up to one day below the threshold allowed within an event. Values shown next to the legend indicate the quasi-stationary wave persistence during 11 December 2024 – 7 January 2025 and the standard deviation of persistence during 1979–2024. (B) Same as (A), but for the distribution of quasi-stationary wave persistence in NeuralGCM large ensemble (gray), involution ensemble (red), and AMIP-ClimSST experiments (blue).

To test this hypothesis, we quantify quasi-stationary wave persistence as the number of consecutive days with a strong quasi-stationary wave, defined by area-mean eddy geopotential height anomalies over the western United States exceeding one temporal standard deviation. We retain only events lasting at least three days, while allowing for up to one day below the threshold within an event. Figure 6A shows the quasi-stationary wave persistence during 11 December 2024 – 7 January 2025 in ERA5, together with its distribution over the past 46 years. During December 2024/January 2025, the strong quasi-stationary wave persisted for 11 days, which is substantially longer than both the long-term median (6 days) and the standard deviation (6.2 days) during 1979–2024.

Both the NeuralGCM large ensemble and the involution ensemble reproduce this prolonged quasi-stationary wave persistence, with mean durations of 10.9 and 10.4 days, respectively (Fig. 6B). In contrast, the AMIP-ClimSST experiment, which excludes the observed La Niña SST anomalies, yields a shorter persistence of 8.4 days, significantly less than that of the involution ensemble ($p < 0.06$). These results indicate that the observed La Niña SST anomalies played an important role prolonging the quasi-stationary wave over the western United States, even though atmospheric initial conditions were the dominant driver of its occurrence.

## 4. Conclusion

In December 2024, a pronounced quasi-stationary anticyclonic circulation persisted over the western United States for several weeks before the January 2025 LA wildfires. Because predicting the large-scale circulation patterns that favor fire-conducive conditions is of great practical importance, we developed an involution-ensemble framework based on ML-based large-ensemble simulations to improve subseasonal predictability of extreme events such as the January 2025 LA Wildfires. Our main conclusions are as follows:

1) NeuralGCM involution ensemble skillfully reproduces the temporal evolution of the quasi-stationary wave over the western United States during December 2024 and early January 2025, despite being constrained only during the 5–10 December 2024. The involution ensemble captures the observed evolution of the quasi-stationary wave with a temporal correlation of 0.56 ($p < 0.001$). Sensitivity experiments using the AMIP-IC and AMIP-ClimSST configurations further show that atmospheric initial conditions are the dominant source of quasi-stationary wave predictability, accounting for about 52% of the total quasi-stationary wave variation leading up to the wildfires.

2) The occurrence of the quasi-stationary wave is closely linked to active MJO conditions in phases 6–7, while background La Niña SST anomalies help prolong its persistence. Specifically, during active MJO phase 6–7 and under La Niña-like SST anomalies, anomalous convection over the tropical western Pacific excites a Rossby wave train that propagates along the subtropical jet and favors downstream anticyclonic circulation over the western United States. Strong quasi-stationary wave events occur with a probability of about 31% during the nine days following active MJO phases 6–7, about 35% higher than during other active MJO phases (23% [21%~25%]). Besides, the observed La Niña SST anomalies are associated with a quasi-stationary wave persistence of 11 days, compared with a climatological median persistence of 6 days.
3) The NeuralGCM involution ensemble reasonably captures both the MJO evolution and La Niña-related enhancement of quasi-stationary wave persistence. During the prediction period beginning 11th December 2024, about 37% of days in the 1000-member large ensemble fall within MJO phases 6–7. By comparison, the 50-member involution ensemble yields a slightly higher fraction, about 41%, indicating that the involution workflow effectively identifies members with enhanced predictive skill. Besides, the large and involution ensembles simulate quasi-stationary wave persistence of 10.9 and 10.4 days, respectively, both longer than the 8.4 days simulated in the climatological-SST experiment.

In addition to these large-scale drivers, Santa Ana winds are well known to strongly influence wildfire risk in Southern California, including during the January 2025 LA wildfires (Westerling et al. 2004; Hughes and Hall 2010; Lindsey 2025; Srivastava et al. 2026). A natural follow-up question is whether the NeuralGCM involution ensemble can also capture the local thermodynamic and dynamical conditions associated with Santa Ana events. To examine this, following Hughes and Hall (2010), we define a katabatic pressure-gradient index and a Santa Ana wind-direction index, both of which quantify the thermodynamic contrast between the inland topography northwest of LA and the adjacent offshore ocean region (see Appendix). Positive values of either index indicate enhanced offshore flow from inland Southern California toward the coast. The results suggest that the anomalous Santa Ana winds are linked to the strong quasi-stationary wave preceding the LA wildfires in the NeuralGCM involution ensemble during 11 December 2024 – 8 January 2025, as indicated by a significant positive correlation between anomalous eddy geopotential height over the western United States and the Santa Ana wind-direction index (Figs. S3A & S3B). Besides, large positive katabatic pressure gradients tend to coincide with positive eddy geopotential height anomalies and positive wind-direction indices (red shading for the dots). A similar relationship is found in the AMIP-ClimSST experiment with realistic atmospheric initial conditions (Fig. S3C).

Taken together, these results demonstrate the subseasonal predictability of persistent large-scale circulation patterns preceding extreme wildfires by up to about one month. Recent studies have extended atmospheric predictability beyond conventional limits by applying ML-based optimization or fine tuning to model structure or atmospheric initial conditions (Kochkov et al. 2024; Duan et al. 2025; Mahesh et al. 2025a,b; Vonich and Hakim 2026). At the same time, prediction skill can also be improved by deepening our understanding of the physical sources of predictability (e.g., Toms et al. 2020; Hakim and Masanam 2024; Shin et al. 2024; Zhang et al. 2025a; Chen et al. 2026; Hassanzadeh et al. 2026; Xiang et al. 2026). As recent advances in ML have greatly reduced the computational cost of large-ensemble forecasting (Henn et al. 2024; Duan et al. 2025; Kent et al. 2025; Chen et al. 2026), an important challenge is how to make the most effective use of these large ensembles. The involution-ensemble framework provides one such pathway by combining ML-based large-ensemble predictions with physics-guided ensemble sampling to identify windows of opportunity for anticipating low-likelihood high-impact extreme events on subseasonal timescales.

More broadly, the involution-ensemble framework may also be useful for regional refinement and storyline-based assessment. Here we apply the involution ensemble workflow over the broader North America domain, yet it still reproduced the temporal evolution of the large-scale circulation pattern over the western United States with reasonable skill. This suggests that the framework may be particularly valuable for regional S2S prediction in the domains with dense and reliable observations. By capturing the dynamical evolution of extreme events, it may also support process-oriented attribution and the development of physically grounded storylines for low-likelihood high-impact events in a changing climate.

*Acknowledgments.*

This study is supported by Office of Science, U.S. Department of Energy (DOE) Biological and Environmental Research as part of the Regional and Global Model Analysis program area through the Water Cycle: Modeling of Circulation, Convection, and Earth System Mechanisms (WACCEM) scientific focus area. The Pacific Northwest National Laboratory (PNNL) is operated for DOE by Battelle Memorial Institute under contract DE-AC05-76RL01830. We acknowledge Dr. Stephan Hoyer, Dr. Janni Yuval, and Dr. Dmitrii Kochkov from Google Research, and Dr. Letian Gu for the help and constructive suggestions to improve the NeuralGCM long-term free- running simulations. We also acknowledge Karthik Balaguru for the guidance of manuscript proposal submitted to BAMS.

*Data Availability Statement.*

All data needed to evaluate the conclusions in the paper are presented in the paper. The relevant codes and data in this work are available, upon request, from the corresponding author Z. C. and L. R. L. The US Government retains and the publisher, by accepting this article for publication, acknowledges that the US Government retains a non-exclusive, paid-up, irrevocable, world-wide license to publish or reproduce the published form of this manuscript, or allow others to do so for US Government purposes. The Department of Energy will provide public access to these results of federally sponsored research in accordance with the DOE Public Access Plan: (http://energy.gov/downloads/doe- public-access-plan).

## APPENDIX

### MJO index and event definition

The near-real-time observed MJO index is obtained from the Australian Bureau of Meteorology. The index is calculated following Wheeler and Hendon (2004) through 2013, with the revised methodology of Gottschalck et al. (2010) applied from 2014 onward. It is derived from the leading two EOFs of the combined fields of outgoing longwave radiation OLR, 850-hPa zonal wind (U850), and 200-hPa zonal wind (U200), averaged over 15ºS–15ºN. Before the EOF analysis, the time mean and the first three harmonics of the annual cycle are removed from each field, followed by subtraction of the trailing 120-day mean to isolate intraseasonal variability. The resulting anomaly fields are then averaged over 15ºS–15ºN, and each variable is normalized by its global variance prior to the combined EOF analysis. T processed daily fields are projected onto the fixed leading two EOFs to obtain the real-time multivariate MJO indices, RMM1 and RMM2.

$$RMM_k(t) = \boldsymbol{EOF}_k^T \times \boldsymbol{x}(t), \text{ (k = 1, 2)}, \quad (1)$$

where $\boldsymbol{x}(t)$ is the vector containing the preprocessed and normalized OLR, U850 and U200 fields at longitude grid points, and $\boldsymbol{EOF}_k$ is the k^th multivariate EOF mode, consisting of three components corresponding to OLR, U850, and U200. The MJO amplitude is then calculated as $\sqrt{RMM1^2 + RMM2^2}$. The MJO is classified as active if its amplitude is greater than 1. The two-dimensional phase space defined by RMM1 and RMM2 is divided into eight equal sectors, corresponding to the eight canonical MJO phases.

When calculating the RMM indices in NeuralGCM simulations, a modified methodology is required because NeuralGCM does not output OLR:

$$RMM_{k,NGCM}(t) = \boldsymbol{EOF}_{k,U850,U200}^{T} \times \binom{\boldsymbol{U850}}{\boldsymbol{U200}}(t),\ (\mathrm{k} = 1, 2), \tag{2}$$

where $\boldsymbol{EOF}_{k,U850,U200}^{T}$ is the kth multivariate EOF mode derived from observations, but retaining only the two circulation components. $\binom{\boldsymbol{U850}}{\boldsymbol{U200}}(t)$ represents the matrix of circulation fields simulated by NeuralGCM.

## Estimation of conditional probabilities

To examine the influence of MJO-related remote forcing on quasi-stationary wave occurrence, we calculate the conditional probability (or frequency) of strong quasi-stationary wave (QSW) events following active MJO phases, as shown in Figure 5A, using ERA5 data in 1979–2025:

$$P_{l,g} = P(QSW^*(t + l) \geq 0 | MJO(t) \in g, A(t) > 1). \tag{3}$$

Strong quasi-stationary wave days are defined as days when the standardized area-mean 500-hPa eddy geopotential height anomaly over the western United States exceeds 0:

$$QSW^*(t) = \frac{QSW(t) - QSW_{Dec2024}}{\sigma_{QSW,Dec-Jan}} \geq 0, \tag{4}$$

where g denotes the MJO phase pairs. To increase sample size, adjacent MJO phases with similar convective structures are combined into four phase pairs: 8–1, 2–3, 4–5, and 6–7 (see Fig. 5). The set $l$ denotes the daily lag window considered in the calculation, and A is the MJO amplitude. Only active MJO events with A>1 are included.

Similarly, to identify which MJO phase pairs most often precede strong quasi-stationary wave events, we calculate the conditional probability shown in Figure 5B as:

$$P_{p,g} = P(MJO(t - l) \in g, A(t - l) > 1 \mid QSW^*(t) \geq 0),\ \ l \in l_p, \tag{5}$$

## Logistic regression model

To further quantify occurrence of strong quasi-stationary wave events following specific active MJO phases, we construct a logistic regression model with phase-pair indicators using ERA5 data for 1979–2025:

$$log\left[\frac{P(QSW^*(t+l)\geq 0 \mid I_g(t)=1)}{1-P(QSW^*(t+l)\geq 0 \mid I_g(t)=1)}\right] = \alpha_l + \beta_{g,l} I_g(t), \tag{6}$$

where $I_g(t) = \begin{cases} 1, & MJO(t) \in g \; and \; A(t) > 1 \\ 0, & MJO(t) \notin g \; and \; A(t) > 1 \end{cases}$ is a binary variable indicating the occurrence of a strong quasit-stationary wave event. The corresponding odds ratio (OR) is then calculated as:

$$OR_{g,l} = exp(\beta_{g,l}). \tag{7}$$

An OR>1 indicates that the corresponding MJO phase pair is associated with a higher probability of strong quasi-stationary wave occurrence than all other active phase pairs or then all active MJO days combined, and vice versa. Take MJO phase pair 6–7 as an example, let $P_{p6-7} = P(QSW^*(t+l) \geq 0 \mid I_g(t) = 1)$:

$$P_{P6-7}/(1 - P_{P6-7}) = exp(\alpha + \beta). \tag{8}$$

Then the probability of strong quasi-stationary wave occurrence is given by:

$$P_{P6-7} = exp(\alpha + \beta)/[1 + exp(\alpha + \beta)]. \tag{9}$$

## Santa Ana wind direction index and katabatic pressure gradient

To assess whether the NeuralGCM involution ensemble captures the Santa Ana winds and the associated local thermodynamic forcing during the January 2025 LA wildfires, we calculate a katabatic pressure gradient and define a Santa Ana wind-direction index, following Hughes and Hall (2010). The katabatic pressure gradient is defined as:

$$\mathcal{B} = \frac{g\theta'}{\theta_0} sin(\alpha), \tag{9}$$

where $\theta'$ indicates the temperature deficit of the cold layer between southwestern ocean domain (32ºN~34ºN, 119ºW~121ºW) and the northeastern desert surface domain (34ºN~36ºN, 117ºW~119ºW) at the same altitude near LA. $\theta_0$ denotes the mean surface air temperature over the northeastern desert domain. $\alpha$ represents the mean topographic slope, which is set to 1º.

The wind-direction index is defined as:

$$WDI = -\vec{u}_{NE} \cdot \vec{u}_{SW} \cdot (u_{NE}/|u_{NE}|), \tag{10}$$

where $\vec{u}_{NE}$ and $\vec{u}_{SW}$ represent the wind vectors over the northeastern desert and southwestern ocean domains, respectively. A wind-direction index greater than 0 indicates offshore flow between two domains, whereas a value less than 0 indicates different wind direction between two domains.

## REFERENCES

Barnes, C., and Coauthors, 2025: *Climate change increased the likelihood of wildfire disaster in highly exposed Los Angeles area*. 1–59pp., https://www.worldweatherattribution.org/wp-content/uploads/WWA-scientific-report-LA-wildfires-1.pdf.

Cazzaniga, G., and D. Faranda, 2025: January 2025 California wildfires have been fueled by meteorological conditions strengthened by human-driven climate change.

Chen, Y., D. C. Morton, N. Andela, L. Giglio, and J. T. Randerson, 2016: How much global burned area can be forecast on seasonal time scales using sea surface temperatures ? How much global burned area can be forecast on seasonal time scales using sea surface temperatures ? *Environ. Res. Lett.*, **11**, 045001, https://doi.org/10.1088/1748-9326/11/4/045001.

Chen, Z., J. Lu, C. C. Chang, S. W. Lubis, and L. R. Leung, 2023: Projected increase in summer heat-dome-like stationary waves over Northwestern North America. *npj Clim. Atmos. Sci.*, **6**, 1–10, https://doi.org/10.1038/s41612-023-00511-2.

——, S. W. Lubis, J. Lu, C. C. Chang, H. Huang, W. Zhou, B. E. Harrop, and L. R. Leung, 2025: Eastward-shifting boreal summer intraseasonal oscillation amplifies North America heatwave and wildfire risks in warming climate. *npj Clim. Atmos. Sci.*, **8**, 1–14, https://doi.org/10.1038/s41612-025-01196-5.

——, and Coauthors, 2026: Hierarchical Testing of a Hybrid Machine Learning-Physics Global Atmosphere Model. *AGU Adv.*, **7**, 1–23, https://doi.org/10.1029/2025AV002075.

Diao, C., and E. A. Barnes, 2025: Assessing MJO Tropical-Extratropical Teleconnections in Deep Learning Weather Prediction Models.

Dong, L., L. R. Leung, Y. Qian, Y. Zou, F. Song, and X. Chen, 2021: Meteorological Environments Associated With California Wildfires and Their Potential Roles in Wildfire Changes During 1984 – 2017 Journal of Geophysical Research : Atmospheres. *J. Geophys. Res. Atmos.*, **126**, e2020JD033180., https://doi.org/10.1029/2020JD033180.

Duan, S., J. Zhang, C. Bonfils, and G. Pallotta, 2025: Testing NeuralGCM's capability to simulate future heatwaves based on the 2021 Pacific Northwest heatwave event. *npj Clim. Atmos. Sci.*, **8**, 1–8, https://doi.org/10.1038/s41612-025-01137-2.

Fernandes, L., and A. M. Grimm, 2023: ENSO Modulation of Global MJO and Its Impacts on South America. *J. Clim.*, **36**, 7715–7738, https://doi.org/10.1175/JCLI-D-22-0781.1.

Forkel, M., W. Dorigo, G. Lasslop, I. Teubner, E. Chuvieco, and K. Thonicke, 2017: A data-driven approach to identify controls on global fire activity from satellite and climate observations (SOFIA V1). *Geosci. Model Dev.*, **10**, 4443–4476, https://doi.org/10.5194/gmd-10-4443-2017.

Forkel, M., and Coauthors, 2019: Emergent relationships with respect to burned area in global satellite observations and fire-enabled vegetation models. *Biogeosciences*, **16**, 57–76, https://doi.org/10.5194/bg-16-57-2019.

Gottschalck, J., and Coauthors, 2010: A Framework for Assessing Operational Madden–Julian Oscillation Forecasts. *Bull. Am. Meteorol. Soc.*, **91**, 1247–1258, https://doi.org/10.1175/2010BAMS2816.1.

Hakim, G. J., and S. Masanam, 2024: Dynamical Tests of a Deep Learning Weather Prediction Model. *Artif. Intell. Earth Syst.*, **3**, 1–11, https://doi.org/10.1175/aies-d-23-0090.1.

Hassanzadeh, P., W. Li, Y. Q. Sun, J. Wang, A. Wikner, J. Finkel, and J. Q. Weare, 2026: Missing the Butterfly and Predicting the Past : Features or Bugs of Accurate AI Weather Models? 1–45.

Henderson, S. A., and E. D. Maloney, 2018: The Impact of the Madden-Jullian Oscillation on High-Latitude Winter Blocking during El Nino-Southern Oscillation Events. *J. Clim.*, **31**, 5293–5318, https://doi.org/10.1175/JCLI-D-17-0721.1.

——, ——, and E. A. Barnes, 2016: The Influence of the Madden – Julian Oscillation on Northern Hemisphere Winter Blocking. *J. Clim.*, **29**, 4597–4616, https://doi.org/10.1175/JCLI-D-15-0502.1.

——, ——, and S. W. Son, 2017: Madden-Julian oscillation Pacific teleconnections: The impact of the basic state and MJO representation in general circulation models. *J. Clim.*, **30**, 4567–4587, https://doi.org/10.1175/JCLI-D-16-0789.1.

Henn, B., and Coauthors, 2024: AIMIP Phase 1 : systematic evaluations of AI weather and climate models.

Hersbach, H., 2000: Decomposition of the continuous ranked probability score for ensemble

prediction systems. *Weather Forecast.*, **15**, 559–570, https://doi.org/10.1175/1520-0434(2000)015<0559:DOTCRP>2.0.CO;2.

Huang, B., and Coauthors, 2017: Extended reconstructed Sea surface temperature, Version 5 (ERSSTv5): Upgrades, validations, and intercomparisons. *J. Clim.*, **30**, 8179–8205, https://doi.org/10.1175/JCLI-D-16-0836.1.

Hughes, M., and A. Hall, 2010: Local and synoptic mechanisms causing Southern California ’ s Santa Ana winds. *Clim. Dyn.*, **34**, 847–857, https://doi.org/10.1007/s00382-009-0650-4.

Jones, M. W., and Coauthors, 2022: Global and Regional Trends and Drivers of Fire Under Climate Change. *Rev. Geophys.*, **60**, 1–76, https://doi.org/10.1029/2020RG000726.

Justino, F., D. H. Bromwich, V. Schumacher, A. DaSilva, and S. Wang, 2020: Arctic Oscillation and Pacific-North American pattern dominated-modulation of fire danger and wild fire occurrence. *npj Clim. Atmos. Sci.*, **5**, 1–13, https://doi.org/10.1038/s41612-022-00274-2.

Kautz, L. A., O. Martius, S. Pfahl, J. G. Pinto, A. M. Ramos, P. M. Sousa, and T. Woollings, 2022: Atmospheric blocking and weather extremes over the Euro-Atlantic sector - A review. *Weather Clim. Dyn.*, **3**, 305–336, https://doi.org/10.5194/wcd-3-305-2022.

Kent, C., A. A. Scaife, N. J. Dunstone, D. Smith, S. C. Hardiman, and O. Watt-meyer, 2025: Skilful global seasonal predictions from a machine learning weather model trained on reanalysis data. *npj Clim. Atmos. Sci.*, **8**, 1–17, https://doi.org/10.1038/s41612-025-01198-3.

Kochkov, D., and Coauthors, 2024: Neural general circulation models for weather and climate. *Nature*, https://doi.org/10.1038/s41586-024-07744-y.

LAEDC, 2025: *IMPACT OF 2025 LOS ANGELES WILDFIRES AND COMPARATIVE STUDY*. 1–13pp., https://laedc.org/wp-content/uploads/2025/02/LAEDC_2025-LA-Wildfires-Study_090525-UPDATE.pdf.

Lee, R. W., S. J. Woolnough, A. J. Charton-Perez, and F. Vitart, 2019: ENSO Modulation of MJO Teleconnections to the North Atlantic and Europe Geophysical Research Letters. *Geophys. Res. Lett.*, **46**, 13535–13545, https://doi.org/10.1029/2019GL084683.

Lindsey, R., 2025: The weather and climate influences on the January 2025 fires around Los

Angeles.

Liu, S., S. Hu, and R. Seager, 2024a: The West Pacific Teleconnection Drives the Interannual Variability of Autumn Wildfire Weather in the Western United States After 2000. *Earth's Futur.*, **12**, e2024EF004922, https://doi.org/10.1029/2024EF004922.

Liu, Y., Y. Qian, P. J. Rasch, K. Zhang, L. R. Leung, and Y. Wang, 2024b: Fire – precipitation interactions amplify the quasi-biennial variability in fires over southern Mexico and Central America. *Atmos. Chem. Phys.*, **24**, 3115–3128, https://doi.org/10.5194/acp-24-3115-2024.

Lubis, S. W., Z. Chen, J. Lu, S. Hagos, C.-C. Chang, and L. R. Leung, 2024: Enhanced Pacific Northwest Heat Extremes and Wildfire Risks Induced by the Boreal Summer Intraseasonal Oscillation. *npj Clim. Atmos. Sci.*, **7**, 1–15, https://doi.org/10.1038/s41612-024-00766-3.

Ma, Q., L. Wei, Y. Wang, G. J. Zhang, X. Zhou, and B. Wang, 2025: Fire heat affects the impacts of wildfires on air pollution in the United States. *Science (80-. ).*, **389**, 1137–1143, https://doi.org/10.1126/science.ads1957.

Madakumbura, G., C. Thackeray, A. Hall, P. Williams, J. Norris, and R. Sukhdeo, 2025: Study: Climate Change A Factor In Unprecedented LA Fires. *UCLA*. Accessed 28 October 2026, https://sustainablela.ucla.edu/2025lawildfires.

Mahesh, A., and Coauthors, 2025a: Huge ensembles – Part 1 : Design of ensemble weather forecasts using spherical Fourier neural operators. *Geosci. Model. Dev*, **18**, 5575–5603, https://doi.org/10.5194/gmd-18-5575-2025.

——, and Coauthors, 2025b: Huge ensembles – Part 2 : Properties of a huge ensemble of hindcasts generated with spherical Fourier neural operators. *Geosci. Model Dev.*, **18**, 5605–5633, https://doi.org/10.5194/gmd-18-5605-2025.

Mori, M., and M. Watanabe, 2008: The Growth and Triggering Mechanisms of the PNA : A MJO – PNA Coherence. *J. Meteorol. Soc. Japan*, **86**, 213–236, https://doi.org/10.2151/jmsj.86.213.

Peings, Y., C. Dong, A. Mahesh, M. S. Pritchard, W. D. Collins, and G. Magnusdottir, 2025: Subseasonal forecasting and MJO teleconnections in machine learning weather prediction models.

Qian, Y., 2025: Burning questions on wildfire. *Science (80-. ).*, **389**, 1086–1087, https://doi.org/10.1126/science.aea7430.

Rolla, J., K. Miner, N. Spangler, C. Elder, E. Howard, and J. Milton, 2026: Remote Sensing During the 2025 Los Angeles Wildfires. *AGU Adv.*, **7**, e2025AV002092, https://doi.org/10.1029/2025AV002092.

Roundy, P. E., K. MacRitchie, J. Asuma, and T. Melino, 2010: Modulation of the Global Atmospheric Circulation by Combined Activity in the Madden–Julian Oscillation and the El Nino–Southern Oscillation during Boreal Winter. *J. Clim.*, **23**, 4045–4059, https://doi.org/10.1175/2010JCLI3446.1.

Selz, T., and G. C. Craig, 2026: Can AI-Based Weather Prediction Models Simulate the Butterfly Effect? The Role of Architecture and Implementation. *J. Geophys. Res. Mach. Learn. Comput.*, **3**, e2025JH001180, https://doi.org/10.1029/2025JH001180.

Shin, N. Y., D. Kim, D. Kang, H. Kim, and J. S. Kug, 2024: Deep learning reveals moisture as the primary predictability source of MJO. *npj Clim. Atmos. Sci.*, **7**, 1–8, https://doi.org/10.1038/s41612-023-00561-6.

Simmons, A. J., J. M. Wallace, and G. Branstator, 1983: Barotropic Wave Propagation and Instability, and Atmospheric Teleconnection Patterns. *J. Atmos. Sci.*, **40**, 1361–1392, https://doi.org/10.1175/1520-0469(1983)040<1363:BWPAIA>2.0.CO;2.

Srivastava, A., M. Risser, P. A. Ullrich, and M. Wehner, 2026: Influence of natural and anthropogenic forcings on western US winter fire weather and the 2025 Los Angeles wildfires. *Environ. Res. Lett.*, https://doi.org/10.1088/1748- 9326/aeb710.

Stan, C., D. M. Straus, J. S. Frederiksen, H. Lin, E. D. Maloney, and C. Schumacher, 2017: Review of tropical-extratropical teleconnections on intraseasonal time scales. *Rev. Geophys.*, **55**, 902–937, https://doi.org/10.1002/2016RG000538.

Ting, M., and L. Yu, 1998: Steady response to tropical heating in wavy linear and nonlinear baroclinic models. *J. Atmos. Sci.*, **55**, 3565–3582, https://doi.org/10.1175/1520-0469(1998)055<3565:SRTTHI>2.0.CO;2.

Toms, B. A., E. A. Barnes, and I. Ebert-Uphoff, 2020: Physically Interpretable Neural Networks for the Geosciences: Applications to Earth System Variability. *J. Adv. Model. Earth Syst.*, **12**, 1–20, https://doi.org/10.1029/2019MS002002.

Vonich, P. T., and G. J. Hakim, 2024: Predictability Limit of the 2021 Pacific Northwest Heatwave From Deep-Learning Sensitivity Analysis. *Geophys. Res. Lett.*, **51**, e2024GL110651., https://doi.org/10.1029/2024GL110651.

——, and ——, 2026: Atmospheric Predictability beyond 30 Days with Machine Learning. *Artif. Intell. Earth Syst.*, **5**, 1–16, https://doi.org/10.1175/AIES-D-26-0009.1.

Wallace, J. M., and D. S. Gutzler, 1981: Teleconnections in the Geopotential Height Field during the Northern Hemisphere Winter. *Mon. Weather Rev.*, **109**, 784–812, https://doi.org/10.1175/1520-0493(1981)109<0784:TITGHF>2.0.CO;2.

Wallace, J. M., C. Ng, D. S. Battisti, Q. Fu, J. E. Kim, K. S. Virts, and L. R. Leung, 2026: The Increasing Incidence of Boreal and Western U.S. Forest Wildfires and Its Causes: An Overview. *Bull. Am. Meteorol. Soc.*, **107**, 2112–2129, https://doi.org/10.1175/BAMS-D-25-0193.1.

Wang, J., M. J. Deflorio, A. Gershunov, F. M. Ralph, K. Guirguis, and L. D. Monache, 2024: Association of western US compound hydrometeorological extremes with Madden-Julian oscillation and ENSO interaction. *Commun. Earth Environ.*, **5**, 1–12, https://doi.org/10.1038/s43247-024-01449-w.

Wang, S. S., Y. Qian, L. R. Leung, and Y. Zhang, 2022: Interpreting machine learning prediction of fire emissions and comparison with FireMIP process-based models. *Atmos. Chem. Phys.*, **22**, 3445–3468, https://doi.org/10.5194/acp-22-3445-2022.

Wang, S. S. C., L. R. Leung, Y. Qian, and R. Grotjahn, 2026: Predicting and Explaining Lightning and Lightning - Ignited Wildfires in Northern California With a Two - Step Convolutional Neural Network Framework. *J. Geophys. Res. Atmos.*, **131**, e2026JD046634., https://doi.org/10.1029/2026JD046634.

Ward-Baranyay, M., C. Lee, M. Pascolini-Campbell, D. Sousa, and A. M. Kinoshita, 2026: Pre-Fire Fuel Conditions Are Dominant Drivers of Burn Severity in the 2025 Los Angeles County Fires. *AGU Adv.*, **7**, e2025AV002179, https://doi.org/10.1029/2025AV002179.

Westerling, A. L., D. R. Cayan, T. J. Brown, B. L. Hall, and L. G. Riddle, 2004: Climate , Santa Ana Winds and Autumn Wildfires in Southern California. *Eos (Washington. DC).*, **85**, 1–7, https://doi.org/10.1029/2004EO310001Digital Object Identifier (DOI).

Wheeler, M. C., and H. H. Hendon, 2004: An all-season real-time multivariate MJO index: Development of an index for monitoring and prediction. *Mon. Weather Rev.*, **132**, 1917–1932, https://doi.org/10.1175/1520-0493(2004)132<1917:AARMMI>2.0.CO;2.

WMO, 2025: La Niña event is expected to be short-lived. *Press Release*, 1–5.

Xiang, L., and Coauthors, 2026: Revealing the Mechanisms of Heat Extremes Using an AI Enabled Diagnostic Framework. *Geophys. Res. Lett.*, **53**, e2025GL120917, https://doi.org/10.1029/2025GL120917.

Yuval, J., I. Langmore, D. Kochkov, and S. Hoyer, 2026: Neural general circulation models optimized to predict satellite-based precipitation observations. *Sci. Adv.*, **12**, 1–11, https://doi.org/10.1126/sciadv.adv6891.

Zhang, G., M. Rao, J. Yuval, and M. Zhao, 2025a: Advancing seasonal prediction of tropical cyclone activity with a hybrid AI-physics climate model. *Environ. Res. Lett.*, **20**, 1–11, https://doi.org/10.1088/1748-9326/adf864.

Zhang, W., S. S. Wang, Y. Chikamoto, R. Gillies, and M. Laplante, 2025b: A weather pattern responsible for increasing wildfires in the western United States. *Environ. Res. Lett.*, **20**, 014007, https://doi.org/10.1088/1748-9326/ad928f.

Zhou, W., D. Yang, S. P. Xie, and J. Ma, 2020: Amplified Madden–Julian oscillation impacts in the Pacific–North America region. *Nat. Clim. Chang.*, **10**, 654–660, https://doi.org/10.1038/s41558-020-0814-0.